\documentclass[11pt,a4paper,leqno]{article}

\usepackage{latexsym}
\usepackage{amssymb}

\newif\ifpdf
\ifx\pdfoutput\undefined
   \pdffalse
\else
   \pdfoutput=1
   \pdftrue
\fi

\ifpdf
   \usepackage[pdftex,
               pdftitle={Expected Shortfall: a natural coherent alternative to Value at Risk},
               pdfsubject={How to create a coherent version of Expected Shortfall?},
               pdfkeywords={Expected Shortfall; Risk measure; worst conditional expectation; tail conditional expectation;
               value-at-risk (VaR); conditional value-at-risk (CVaR); tail mean; coherence; quantile;
               sub-additivity},
               pdfauthor={Carlo Acerbi, Dirk Tasche},
               pdfstartview=FitBH,
               pdfview=FitBH,
               colorlinks=true]{hyperref}
\else
\fi

\newtheorem{theorem}{Theorem}

\newtheorem{definition}[theorem]{Definition}

\title{Expected Shortfall: a natural coherent alternative \\ to Value at Risk}

\author{
Carlo Acerbi\thanks{Abaxbank, Corso Monforte 34, 20122 Milano, Italy;
E-mail: carlo.acerbi@abaxbank.com}
\quad
Dirk Tasche\thanks{Zentrum Mathematik (SCA), TU M\"unchen, 80290 M\"unchen,
Germany; E-mail: tasche@ma.tum.de}
}

\date{May 9, 2001}

\begin{document}
\maketitle
\begin{abstract}
We discuss the coherence properties of Expected Shortfall ($ES$) as a financial risk measure. This statistic arises in a natural way from the estimation of the ``average of the $100p \%$ worst losses'' in a sample of returns to a portfolio. Here $p$ is some fixed confidence level. We also compare several alternative representations of $ES$ which turn out
to be more appropriate for certain purposes.

\end{abstract}


\section{A four years impasse}

Risk professionals have been looking for a coherent alternative to Value at Risk ($\mathrm{VaR}$) for four years. Since the appearance, in 1997, of {\em Thinking Coherently} by Artzner et al \cite{ADEH97} followed by {\em Coherent Measures of Risk} 
\cite{ADEH99}, it was clear to risk practitioners and researchers that the gap between market practice and theoretical progress had suddenly widened enormously. These papers in fact faced for the first time the problem of defining in a clearcut way what properties a statistic of a portfolio should have in order to be considered a sensible risk measure. The answer to this question was given through a complete characterization of such properties via an axiomatic formulation of the concept of {\em coherent risk measure}. With this result, risk management became all of a sudden a science in itself with its own rules correctly defined in a deductive framework. Surprisingly enough, however, $\mathrm{VaR}$, the risk measure adopted as best practice by essentially all banks and regulators, happened to fail the exam for being admitted in this science. $\mathrm{VaR}$ is not a coherent risk measure because it simply doesn't fulfill one of the axioms 
of coherence.

Things are worse than some people strive to believe. The fact that for years the class of coherent measures didn't exhibit any known specimen that shared with $\mathrm{VaR}$ its formidable advantages (simplicity, wide applicability, universality,\ldots) led many practitioners to think that coherence might be some sort of optional property that a risk measure can or cannot display. It seemed that coherent measures belonged to some ideal world which real--world practical risk measures can only dream of. So little attention was paid to this problem that to the best of our knowledge no risk management textbook ever mentioned the fact that $\mathrm{VaR}$ is not coherent.

This attitude means underestimating the impact of the conclusions of \cite{ADEH97}. Writing axioms means crystallizing in a minimal number of precise statements the intrinsic nature of a concept. It is a necessary step to take in the process of translating a complex reality into a mathematical formulation. The axioms of coherence simply embody in a synthetic and essential way those features that single out a risk measure in the class of statistics of a portfolio dynamics, just like the axiom ``it must be higher when air is hotter'' identifies a measure of temperature out of the class of thermodynamical properties of the atmosphere. If you want to use a barometer for measuring temperature despite the fact that pressure does not satisfy the above axiom, don't be surprised if you happen to be dressed like an Eskimo in a hot cloudy day or to be wearing a swim costume in an icy sunshine. 

Broken axioms always lead to paradoxical, wrong results. And $\mathrm{VaR}$ makes no exception. Once you know which axiom is violated by $\mathrm{VaR}$ it is a child's play to provide examples where the assessment of risks via $\mathrm{VaR}$ is definitely wrong or, in other words, where higher $\mathrm{VaR}$ figures come from less risky portfolios \cite{ANS}. 

In this paper and henceforth we are going to take these axioms seriously as many other groups of researchers \cite{ADEH99,BLS00,Em00,Pf00,T00,Ur00}, practitioners and regulators \cite{BoJ} have begun to do. To avoid confusion, if a measure is not coherent we just choose not to call it a risk measure at all. In other words, for us, the above--mentioned axioms define the concept of risk itself via the characterization of the possible operative ways to measure it\footnote{Note that this was indeed the genuine motivation of \cite{ADEH99}. Quoting from the introduction: {\em We provide in this paper a definition of risk \ldots and provide and justify a unified framework for the analysis, construction and implementation of measures of risk}.}. This might seem a dogmatic approach but it is not. We are of course prepared to give up this definition as soon as a new different set of axioms is proposed which is more suitable to a mathematical formulation of the concept of risk measure. What we are not prepared to do anymore, after we learned the lesson of \cite{ADEH97,ADEH99}, is discussing of risk measures without even defining what ``risk measure'' means.

We therefore promote the coherence axioms of \cite{ADEH97} to key defining properties of any risk measure to clearly state that in our opinion speaking of non--coherent measures of risk is as useless and dangerous as speaking of non--coherent measures of temperature. In our language, the adjective coherent is simply redundant.

\begin{definition}[Risk Measure]
\label{def:rm}
Consider a set $V$ of real-valued random variables. 
A function $\rho: V \to \mathbb{R}$ is called a risk measure if it is
\begin{enumerate} 
\item monotonous: $X \in V,\ X \ge 0 \quad \Rightarrow \quad \rho(X) \le 0$,
\item sub--additive: $X, Y, X+Y \in V\quad \Rightarrow \quad \rho(X+Y) \le
\rho(X) + \rho(Y)$,
\item positively homogeneous: $X \in V,\ h > 0, h\,X\in V \quad \Rightarrow \quad \rho(h\,X)
= h\,\rho(X)$, and
\item translation invariant: $X \in V, \ a \in \mathbb{R} \quad \Rightarrow \quad \rho(X + a) =
\rho(X) - a$.
\end{enumerate}  
\end{definition}

$\mathrm{VaR}$ is not a risk measure because it does not fulfill the axiom of sub--additivity. This  property expresses the fact that a portfolio made of sub--portfolios will  risk an amount which is at most the sum of the separate amounts risked by its sub--portfolios. This is maybe the most characterizing feature of a risk measure, something which belongs to everybody's concept of risk. The global risk of a portfolio will be the sum of the risks of its parts only in the case when the latter can be triggered by concurrent events, namely if the sources of these risks may conspire to act altogether. In all other cases, the global risk of the portfolio will be strictly less than the sum of its partial risks thanks to risk diversification. This axiom captures the essence of how a risk measure should behave under the composition/addition of portfolios. It is the key test for checking whether a measurement  of a portfolio's risk is consistent with those of its parts. 

For a sub--additive measure, portfolio diversification always leads to risk reduction, while for measures which violate this axiom, diversification may produce an increase in their value even when partial risks are triggered by mutually exclusive events \cite{ANS}. 

Sub-additivity is necessary for capital adequacy requirements in banking supervision. Think of a bank made of several branches: if the capital requirement of each branch is dimensioned on its own risk, the regulator should be confident that also the overall bank capital should be an adequate one. This may however not be the case if the adopted measure violates sub--additivity since the risk of the whole bank could turn out to be much bigger then the sum of the branches' risks.

Sub-additivity is an essential property also in portfolio--optimization problems. This property in fact is related to the convexity\footnote{In fact convexity follows from sub-additivity and positive homogeneity of definition \ref{def:rm}.} of the risk surface to be minimized in the space of portfolios. Only if the surfaces are convex they will always be endowed with a unique absolute minimum and no fake local minima \cite{Ur00} and the risk minimization process will always pick--up a unique,  well--diversified optimal solution.

Therefore, though one can perfectly think of possible alternative axiomatic definitions of risk measure, we strongly believe that no sensible set of axioms could in any case admit sub-additivity violations.

\section{Constructing a  Risk Measure}

In what follows we want to show how a coherent alternative to Value at Risk arises as the natural answer to  simple questions on a specified sample of worst cases of a distribution. We will construct this measure in a bottom--up fashion to better appreciate that this construction does not leave much freedom and leads in a natural way to essentially one robust solution.

For sake of concreteness, $X$  will be the random variable describing the future value of the profit or loss of a portfolio on some fixed time horizon $T$ from today and $\alpha=A\% \in (0,1)$ will be some percentage which represents a sample of ``worst cases'' for the portfolio that we want to analyze. Provided this information, the $\mathrm{VaR}$ of the portfolio  with parameters $T$ and $A\%$ is simply given by the loss associated with the related  quantile $x^{(\alpha)}$ of the distribution\footnote{We will omit the $T$ dependence where possible.}.
\begin{eqnarray}
        x^{(\alpha)}(X)& =& \sup \{x| \mathrm{P}[X\leq x] \leq \alpha \}\label{eq:1}\\
        \mathrm{VaR}^{(\alpha)}(X) & = & - x^{(\alpha)}(X)\label{eq:2}
\end{eqnarray}
This statistic answers to the following question:
\begin{equation}
\mbox{\em What is the minimum loss incurred in the $A\%$ worst cases of our portfolio?}
\end{equation}
Strange as it may sound, this is the most frequently asked question in financial risk management today. And due to that ``minimum loss'' in its definition $\mathrm{VaR}$ is not a sub--additive measure. Moreover, being simply the threshold of the possible $A\%$ losses, $\mathrm{VaR}$ is indifferent of how serious the losses {\em beyond} that threshold actually are. Little imagination is needed to invent portfolios with identical $\mathrm{VaR}$ and dramatically different levels of risk in the same $A\%$ worst cases sample.

Any reader, at this point is tempted to modify the above question with the following:
\begin{equation} \label{question}
\mbox{\em What is the expected loss incurred in the $A\%$ worst cases of our portfolio?}
\end{equation}
We want to show that this is a  good idea  for at least two different reasons. First of all, because this question is undoubtedly a more natural question to raise when considering the risks of a specified sample of worst cases. Secondly, because it naturally leads to the definition of a sub-additive statistic as we will see in a few steps. 

It is not difficult to understand that if the distribution function of the portfolio is continuous, then the statistic which answers the above question is simply given by a conditional expected value below the quantile or ``tail conditional expectation'' \cite{ADEH97}.
\begin{equation} \label{TCE}
        TCE^{(\alpha)}(X) = - \mathrm{E} \{X |X \leq x^{(\alpha)} \}
\end{equation}
 
For more general distributions however, this statistic does not fit question (\ref{question}) since the event $\{X\leq x^{(\alpha)}\}$ may happen to have a probability larger than $A\%$ and is therefore larger than our set of selected worst cases. Indeed, $TCE$ is a  risk measure only when restricted to continuous distribution functions \cite{T00} while may violate sub-additivity on general distributions \cite{AT01}.

To understand which statistic is actually hidden in question (\ref{question}) let us see how we would naturally answer to it having  a large number $n$ of realizations $\{X_i\}_{\{i=1, \ldots, n\}}$ of the random variable $X$. We simply have to sort the sample in increasing order and average the first $A\%$ values. To do this, define the order statistics $X_{1:n}\leq \ldots\leq X_{n:n}$ as the sorted values of the n--tuple $(X_{1}, \ldots, X_{n})$ and approximate the number of $A\%$ elements in the sample by $w = [n\alpha] = \max\{m|m\leq n\alpha,m\in\mathbb{N}\}$, the  integer part of $n\,A\%$, a choice that for large $n$ could be changed with any other integer rounding or truncation close to $n\alpha$. The set of $A\%$ worst cases is therefore represented by the least $w$ outcomes $\{X_{1:n}, \ldots, X_{w:n}\}$.

Postponing the discussion of some subtleties on quantile estimation we can define the following natural estimator for the $\alpha$--quantile $x^{(\alpha)}$.
\begin{equation}
        x_{n}^{(\alpha)}(X) =  X_{w :n}
\end{equation}
The natural estimator for the expected loss in the $A\%$ worst cases is then simply given by
\begin{equation} \label{ESest}
        ES_{n}^{(\alpha)}(X) = - \frac{\sum_{i=1}^{w}X_{i:n}}{w}
= -\mbox{(Average of least $A\%$ outcomes $X_i$)}
\end{equation}
which we will call the $A\%$ Expected Shortfall of the sample.
Note that then the natural estimator for $TCE$ 
\begin{equation} \label{TCEest}
TCE_{n}^{(\alpha)}(X) =
-\, \frac{\sum_{i=1}^n \, X_i \, {\bf 1}_{\{X_i \leq X_{w:n}\}}}{\sum_{i=1}^n \,  
{\bf 1}_{\{X_i \leq X_{w:n}\}}} = 
-\mbox{(Average of all $X_i\leq x_{n}^{(\alpha)}$)}
\end{equation}
is in general an average of more than $A\%$ of the outcomes\footnote{We adopt the obvious notation 
$\mathbf{1}_{\{\mathrm{Relation}\}} =
\left\{ 
  \begin{array}{c@{\,,\ }l}
1 & \mbox{if Relation is true}\\
0 & \mbox{if Relation is false.}
  \end{array}
\right.
$}. This may happen when the probability of the event $X=x^{(\alpha)}$ is positive (the case of a discrete distribution function) so that there might be multiple occurrences of the  value $X_i=X_{w :n}$. 

It is  easy to see that $ES_{n}^{(\alpha)}$ is indeed sub--additive for any fixed $n$. Consider two variables $X$ and $Y$ and a number $n$ of simultaneous realizations  $\{(X_i,Y_i)\}_{\{i=1, \ldots, n\}}$. We can prove sub-additivity at a glance 
\begin{eqnarray} \label{sub}
        ES_{n}^{(\alpha)}(X+Y) &=& - \frac{\sum_{i=1}^{w}(X+Y)_{i:n}}{w} \\
&\leq& - \frac{\sum_{i=1}^{w}(X_{i:n}+Y_{i:n})}{w} \nonumber \\
&=& ES_{n}^{(\alpha)}(X) + ES_{n}^{(\alpha)}(Y) \nonumber 
\end{eqnarray}
This result is very encouraging. If we understand which statistic $ES_{n}^{(\alpha)}$ is an estimator of for large $n$, we are  likely to end up with a sub--additive measure. Notice, that a proof similar to (\ref{sub}) would fail for $TCE_{n}^{(\alpha)}$.

Now, we can expand the definition of $ES_{n}^{(\alpha)}$
\begin{eqnarray}
ES_{n}^{(\alpha)} &=& -\frac{\sum\limits_{i=1}^{w} X_{i:n} }{w}
= -\frac{\sum\limits_{i=1}^{n} X_{i:n} {\bf 1}_{\{i\leq w\}}}{w}\nonumber \\[2ex]
&=& -
\frac{1}{w}
\left(
\sum_{i=1}^{n} X_{i:n}\, {\bf 1}_{\{X_{i:n}\leq X_{w:n}\}} -
\sum_{i=1}^{n}
X_{i:n}
\Big(
{\bf 1}_{\{X_{i:n}\leq X_{w:n}\}} -
{\bf 1}_{\{i\leq w\}}
\Big)
\right)\nonumber \\[2ex]
&=& -
\frac{1}{w}
\left(
\sum_{i=1}^{n} X_{i}\,{\bf 1}_{\{X_{i}\leq X_{w:n}\}} -
X_{w:n}
\sum_{i=1}^{n}
\Big(
{\bf 1}_{\{X_{i:n}\leq X_{w:n}\}}-
{\bf 1}_{\{i\leq w\}}
\Big)
\right)\nonumber \\[2ex]
&=& -
\frac{n}{w}
\left(\frac{1}{n}
\sum_{i=1}^{n} X_{i}\, {\bf 1}_{\{X_{i}\leq X_{w:n}\}} -
X_{w:n}
\Big(\frac{1}{n}
\sum_{i=1}^{n}
{\bf 1}_{\{X_{i}\leq X_{w:n}\}}-
\frac{w}{n}
\Big)
\right).\label{eq:decomp} 
\end{eqnarray}
If we now had 
\begin{equation}
\lim_{n\to\infty} X_{w:n} \quad = \quad x^{(\alpha)}\,,
  \label{eq:lim}
\end{equation}
with probability 1, it would be easy to conclude that with probability 1 we also have
\begin{equation} 
\lim_{n\to\infty} ES_{n}^{(\alpha)}(X)  = - \frac{1}{\alpha} \Big( \mathrm{E}[ X \,\mathbf{1}_{\{X \le x^{(\alpha)}\}}]
- x^{(\alpha)}\,( \mathrm{P}[X \le x^{(\alpha)}] - \alpha )\Big) 
  \label{eq:lim2}
\end{equation}
Well, this is the subtlety on quantile estimation we have mentioned. Equation (\ref{eq:lim}) does not hold in general. Nevertheless it can be shown \cite{AT01} that eq. (\ref{eq:lim2}) is more robust and in fact holds in full generality. 
We can then give the following
\begin{definition}[Expected Shortfall]
\label{def:es} Let $X$ be the profit--loss of a portfolio on a specified time horizon $T$ and let $\alpha =A\%\in (0,1)$ some specified probability level. The Expected A\% Shortfall of the portfolio is then defined as
\begin{equation}
ES^{(\alpha)}(X) = - \frac{1}{\alpha} \Big( \mathrm{E}[ X \,\mathbf{1}_{\{X \le x^{(\alpha)}\}}]
- x^{(\alpha)}\,( \mathrm{P}[X \le x^{(\alpha)}] - \alpha )\Big)
\label{eq:es}
\end{equation}
\end{definition}
This  definition provides a risk measure perfectly satisfying all the axioms of definition \ref{def:rm}. 
This explicit formulation was first introduced\footnote{This definition may seem different from \cite{ANS} for the use of upper quantile $x^{(\alpha)}$ instead of lower quantile $x_{(\alpha)}$. It can be shown \cite{AT01} that the two formulations actually coincide.} in \cite{ANS} where a general proof of sub-additivity\footnote{It is immediate to verify the other axioms.} was also given which is 
not based on the $n\to\infty$ limit of the above proof (\ref{sub}) of sub-additivity of $ES_{n}^{(\alpha)}$. An implicit formulation of $ES$ had already been proved to be coherent in \cite{Pf00}, where however it was erroneously identified with $TCE$. 

Equation (\ref{eq:es}) might at a first glance look complicated. The concept it expresses is however  simple as it is the literal mathematical translation of  our above natural question and the limit for large $n$ of the straightforward estimator (\ref{ESest}).  It is easy to realize that $TCE$,  despite its simpler mathematical formulation (\ref{TCE}) is on the contrary related to a much more complicated question than (\ref{question}).

To have a better insight of (\ref{eq:es}), the  term $x^{(\alpha)}\,( \mathrm{P}[X \le x^{(\alpha)}] - \alpha )$  has to be interpreted as the exceeding part to be subtracted from the expected value $\mathrm{E}[ X \,\mathbf{1}_{\{X \le x^{(\alpha)}\}}]$ when $\{X \le x^{(\alpha)}\}$ has probability larger than $\alpha=A\%$. When, on the contrary $P[X \le x^{(\alpha)}]=\alpha$, as is always the case if the probability distribution is continuous, the term vanishes and it is easy to see that (\ref{eq:es}) reduces to (\ref{TCE}) or, in other words, $ES^{(\alpha)}=TCE^{(\alpha)}$.

The actual simplicity of $ES^{(\alpha)}$ can be appreciated only giving up defining it as a combination of expected values. There exists in fact an equivalent representation to (\ref{eq:es}) which reveals in a much more transparent way the direct dependence on the parameter $\alpha$ and on the distribution function $F(x)=P[X\leq x]$. In fact, introducing the so--called generalized inverse function of $F(x)$
\begin{equation}
F^\leftarrow(p) = \inf \{x| F(x) \geq p \}
  \label{eq:inv}
\end{equation}
one can easily show \cite{AT01} that $ES^{(\alpha)}$ can be simply expressed as the negative mean of $F^\leftarrow(p)$ on the confidence
level interval $p\in (0,\alpha]$:
\begin{equation}
ES^{(\alpha)}(X) = -\frac{1}{\alpha}\,\int_0^\alpha F^\leftarrow(p) \,dp
  \label{eq:int-es}
\end{equation}
This is the most fundamental formulation of $ES^{(\alpha)}$. Its mathematical tractability makes it particularly appropriate for studying the analytical properties of $ES^{(\alpha)}$. For instance, continuity in $\alpha$ (which is a distinguishing property of $ES^{(\alpha)}$ which  $TCE^{(\alpha)}$ and $VaR^{(\alpha)}$ do not share) is manifest in (\ref{eq:int-es}) while it is not obvious in (\ref{eq:es}).

An alternative useful expression equivalent to (\ref{eq:es}) has been recently formulated in \cite{RU01} where  the terminology\footnote{We prefer the terminology {\em``Expected $A\%$ Shortfall''} because it is more suitable to  the ``expected loss in the $A\%$ worst cases'' while {\em ``$\alpha$--Conditional Value at Risk''} is a name that had been tailored on the conditional expectation of definition (\ref{TCE}).}
 ``$\alpha$--Conditional Value at Risk'' is however adopted for  $ES^{(\alpha)}$
\begin{equation}
ES^{(\alpha)} =  TCE^{(\alpha)} + (\lambda-1)\,(TCE^{(\alpha)}- VaR^{(\alpha)})
\end{equation}
with $\lambda \equiv \mathrm{P}[X \le x^{(\alpha)}]/\alpha \geq 1$. This relationship, which can be easily derived from (\ref{eq:es}) multiplying and dividing by $\mathrm{P}[X \le x^{(\alpha)}]$, allows to put in evidence that in general $ES^{(\alpha)}\geq TCE^{(\alpha)}$. 

\section{Conclusions}

We started with an impasse coming from the fact that $\mathrm{VaR}$ was manifestly shown to be unfit for describing the risks of a portfolio and yet no valid practical alternative was still available in the class of eligible measures of risk.

In this article we have seen that 
at least one specimen of the class of coherent risk measures allows us not to give up any of the advantages people got used to after the advent of $\mathrm{VaR}$. $ES$ is in fact {\em universal}: it can be applied to any instrument and to any underlying source of risk. $ES$ is {\em complete}:  it produces a unique global assessment for portfolios exposed to different sources of risk. $ES$ is (even more than $\mathrm{VaR}$) a {\em simple} concept since it is the answer to a  natural and legitimate question on the risks run by a portfolio.  Furthermore, any bank that has a $\mathrm{VaR}$--based Risk Management system could switch to $ES$ with virtually no additional computational effort. 

Even though a lot of work has still to be done to better investigate the statistical, probabilistic and computational issues raised by the use of ES, we believe that no serious difficulty will be encountered in adapting to it all the techniques developed in recent years for efficient calculations of $\mathrm{VaR}$.


\sloppy


\begin{thebibliography}{99}
\bibitem{ANS} \textsc{Acerbi, C., Nordio, C., Sirtori, C. (2001)} Expected Shortfall as a Tool for Financial Risk Management. Working paper.
\ifpdf
\href{http://www.gloriamundi.org/var/wps.html}
{\tt http://www.gloriamundi.org/var/wps.html}
\else
{\tt http://www.gloriamundi.org/var/wps.html}
\fi
\bibitem{AT01} \textsc{Acerbi, C., Tasche, D. (2001)} On the Coherence of Expected Shortfall. 
Working paper.
\ifpdf
\href{http://www.gloriamundi.org/var/wps.html}
{\tt http://www.gloriamundi.org/var/wps.html}
\else
{\tt http://www.gloriamundi.org/var/wps.html}
\fi
%
\bibitem{ADEH97} \textsc{Artzner, P., Delbaen, F., Eber, J.-M., Heath, D. (1997)} Thinking
coherently. RISK \textbf{10} (11).
%
\bibitem{ADEH99} \textsc{Artzner, P., Delbaen, F., Eber, J.-M., Heath, D. (1999)} 
Coherent measures of risk. Math. Fin. \textbf{9} (3), 203--228.
\bibitem{BLS00} \textsc{Bertsimas, D., Lauprete, G.J., Samarov, A. (2000)}
Shortfall as a risk measure: properties, optimization and applications. 
Working paper, Sloan School of Management, MIT, Cambridge.

\bibitem{Em00} \textsc{Embrechts, P. (2000)} Extreme Value Theory: Potential
and Limitations as an Integrated Risk Management Tool. Working paper, ETH Z\"urich.\\
\ifpdf
\href{http://www.math.ethz.ch/~embrechts/} 
{\tt http://www.math.ethz.ch/$\sim$embrechts/}
\else
\texttt{http://www.math.ethz.ch/$\sim$embrechts/}
\fi

\bibitem{Pf00} \textsc{Pflug, G. (2000)} Some remarks on the value-at-risk
and the conditional value-at-risk. In, Uryasev, S.
(Editor). 2000. Probabilistic Constrained Optimization: Methodology and Applications. Kluwer Academic
Publishers.
\ifpdf
\href{http://www.gloriamundi.org/var/pub.html}
{\tt http://www.gloriamundi.org/var/pub.html}
\else 
\texttt{http://www.gloriamundi.org/var/pub.html}
\fi
%
\bibitem{RU01} \textsc{Rockafellar, R.T., Uryasev, S. (2001)} Conditional Value-at-Risk for general loss distributions. Research report 2001-5, ISE Depart., University of Florida.\\
\ifpdf
\href{http://www.ise.ufl.edu/uryasev/} 
{\tt http://www.ise.ufl.edu/uryasev/}
\else
\texttt{http://www.ise.ufl.edu/uryasev/}
\fi

\bibitem{T00} \textsc{Tasche, D. (2000)} Conditional expectation as
quantile derivative. Working paper, TU M\"unchen.
\ifpdf
\href{http://www.ma.tum.de/stat/}
{\tt http://www.ma.tum.de/stat/}
\else 
\texttt{http://www.ma.tum.de/stat/}
\fi
%
\bibitem{Ur00} \textsc{Uryasev, S. (2000)} Conditional Value-at-Risk: Optimization Algorithms and Applications.
Financial Engineering News \textbf{2} (3).
\ifpdf
\href{http://www.gloriamundi.org/var/pub.html}
{\tt http://www.gloriamundi.org/var/pub.html}
\else 
\texttt{http://www.gloriamundi.org/var/pub.html}
\fi

\bibitem{BoJ} \textsc{Yamai, Y., Yoshiba, T. (2001)} On the validity of 
value-at-risk: comparative analyses with Expected Shortfall.
Discussion paper 2001-E-4, Institute for Monetary and Economic Studies, Bank of Japan.
\ifpdf
\href{http://www.gloriamundi.org/var/wps.html}
{\tt http://www.gloriamundi.org/var/wps.html}
\else
{\tt http://www.gloriamundi.org/var/wps.html}
\fi
\end{thebibliography}
\end{document}